\documentclass[pra,superscriptaddress,twocolumn,dvipsnames]{revtex4}

\usepackage{layout}
\usepackage{color}
\usepackage[utf8]{inputenc}
\usepackage{amsmath}
\usepackage{tikz}
\usepackage{xcolor}
\usetikzlibrary{arrows}
\usetikzlibrary[topaths]

\usepackage{amsfonts}
\usepackage{amsmath}
\usepackage{amsthm}
\usepackage{amscd}
\usepackage{amssymb}
\usepackage{amsxtra}
\usepackage{bm}
\usepackage{bbm}         
\usepackage{exscale}
\usepackage{graphics,color}
\usepackage{latexsym}
\usepackage{enumerate}
\usepackage{dcolumn}      
\usepackage{graphics}
\usepackage{epstopdf}
\usepackage[dvips]{epsfig}
\usepackage{graphicx}
\newcommand{\ket}[1]{\ensuremath{|#1\rangle}}

\newcommand{\ketbra}[2]{\ensuremath{|#1\rangle\langle #2|}}

\newcommand{\beq}{\begin{equation}}
\newcommand{\eeq}{\end{equation}}
\newcommand{\bdf}{\begin{defn}}
\newcommand{\edf}{\end{defn}}

\newcommand{\id}{\ensuremath{\mathbbm 1}}
\newcommand{\tr}{\ensuremath{\text{Tr}}}

\newtheorem{theorem}{Theorem}

\newtheorem{defn}{Definition}
\newtheorem{obs}{Observation}
\newcount\mycount
\begin{document}

\title{\textbf{Practical Quantum Retrieval Games}}

\author{Juan Miguel Arrazola}
\affiliation{Institute for Quantum Computing and Department of Physics and Astronomy, University of Waterloo, 200 University Avenue West, Waterloo, Ontario, N2L 3G1, Canada }
\affiliation{Centre for Quantum Technologies, National University of Singapore, 3 Science Drive 2, Singapore 117543}
\author{Markos Karasamanis}
\affiliation{Department of Physics and Astronomy, University College London,
Gower Street, London WC1E 6BT, United Kingdom}
\author{Norbert L\"{u}tkenhaus}
\affiliation{Institute for Quantum Computing and Department of Physics and Astronomy, University of Waterloo, 200 University Avenue West, Waterloo, Ontario, N2L 3G1, Canada }

\date{\today}

\begin{abstract}
Complex cryptographic protocols are often constructed from simpler building-blocks. In order to advance quantum cryptography, it is important to study practical building-blocks that can be used to develop new protocols. An example is quantum retrieval games (QRGs), which have broad applicability and have already been used to construct quantum money schemes. In this work, we introduce a general construction of quantum retrieval games based on the hidden matching problem and show how they can be implemented in practice using available technology. More precisely, we provide a general method to construct (1-out-of-k) QRGs, proving that their cheating probabilities decrease exponentially in $k$. In particular, we define new QRGs based on coherent states of light, which can be implemented even in the presence of experimental imperfections. Our results constitute a new tool in the arsenal of the practical quantum cryptographer. 
\end{abstract}
\maketitle
In cryptography, the ability to transmit and process quantum information allows us to build cryptographic protocols with properties that are impossible to obtain in a classical setting \cite{bennett84a,wiesner83a,chailloux2009optimal,hillery99a}. Although several such examples are known, exploring new quantum cryptographic protocols and developing the necessary tools to realize them in practice remains an important challenge  \cite{pappa2014experimental,donaldson2016experimental,lunghi2013exp,tang2014experimental,erven2014exp,bash2015quantum}.

Several quantum cryptography protocols, notably those for quantum key distribution \cite{bennett84a}, quantum signature schemes \cite{dunjko2014QDSQKD,arrazola2015multiparty}, oblivious transfer \cite{erven2014exp}, and position-based cryptography \cite{buhrman2014position} , can be constructed from the same basic building-blocks. For instance, it is remarkable that several of these quantum protocols require only that one party is capable of preparing and transmitting the qubit states $\ket{0}, \ket{1},\frac{1}{\sqrt{2}}(\ket{0}+\ket{1})$ and $\frac{1}{\sqrt{2}}(\ket{0}+\ket{1})$, while the receiving party measures them in the corresponding bases. Once this basic building-block is established, more complex protocols can be designed and established. These states, known as the BB84 states \cite{bennett84a}, play a central role in practical quantum cryptography, because preparing and measuring them is a task that can be done routinely in experiments. 

In the effort to advance quantum cryptography and quantum communication, it is important to examine similar \textit{practical} building-blocks that can potentially be used to construct new protocols or even improve existing ones. One example is \emph{quantum retrieval games} (QRGs), which were introduced in Ref. \cite{gavinsky2012quantum} and have been used as building-blocks for quantum money schemes \cite{gavinsky2012quantum,pastawski2012unforgeable,georgiou2015new}. QRGs are a generalization of quantum state discrimination, which is a fundamental problem in quantum information with many applications in cryptography. Therefore, just as with state discrimination, QRGs could potentially be useful in many contexts related to quantum cryptography and quantum communication. Nevertheless, in the general case, QRGs require that the parties prepare complex quantum states of large dimension and perform difficult operations on them, making their implementation challenging.

In this work, we focus on a particular class of QRGs that we call \textit{hidden matching} QRGs, which were first introduced in Ref. \cite{gavinsky2012quantum}. We extend the results of Ref. \cite{gavinsky2012quantum} by considering (1-out-of-k) QRGs and providing a canonical construction for them. We also bound the cheating probability and show that it decays exponentially with $k$. Inspired by this construction, we build a new set of QRGs based on coherent states of light, which can be implemented with available technology. We also examine the role of experimental imperfections such as limited visibility, dark counts and loss, showing that the desired properties of the QRGs can still be attained, especially in the case of large $k$.
\section{Quantum retrieval games}\label{Section: QRGs}
In this section, we review definitions and known results from Refs. \cite{gavinsky2012quantum,georgiou2015new,pastawski2012unforgeable} regarding quantum retrieval games that will be useful for the results presented in this work. 

In a quantum retrieval game (QRG), Alice is given an $n$-bit string $x\in\{0,1\}^n$ selected at random according to some probability distribution $p(x)$. She encodes $x$ into a quantum state $\rho_x$ and sends it to Bob. Bob's goal is to measure $\rho_x$ in order to provide a correct answer to a given question about $x$ with the highest possible probability. Formally, a question about $x$ is modelled as a \emph{relation}. For a set of inputs $X$ and a set of answers $A$, a relation $\sigma$ corresponds to a subset of $X\times A$. Consequently, for a given input $x\in X$ and a relation $\sigma\subseteq X\times A$, we say that an answer $a\in A$ is correct if $(x,a)\in \sigma$. This leads to the following definition of a quantum retrieval game.
\begin{figure}
\begin{center}
\includegraphics[width=0.75\columnwidth]{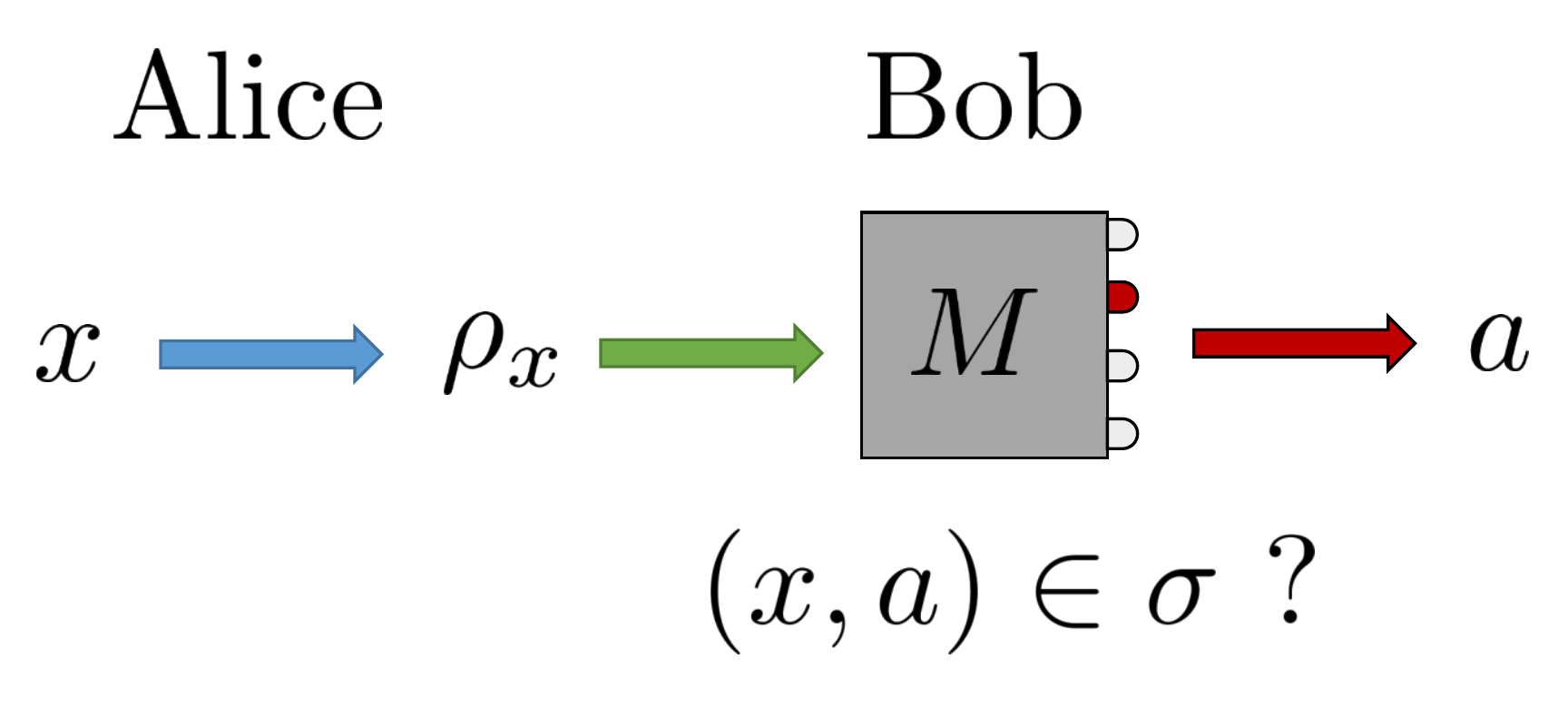}
\caption{A quantum retrieval game. Alice encodes a classical string $x$ into a quantum state $\rho_x$ and sends it to Bob, who performs a measurement $M$ and obtains an outcome $a$. Bob's goal is to obtain an outcome $a$ such that $(x,a)\in\sigma$ with the highest probability possible.}\label{Fig:diagram}
\end{center}
\end{figure}
\begin{defn}
Let $X\subseteq \mathbb{N}$ and $A\subseteq \mathbb{N}$ be respectively the set of inputs and answers. Let also $\sigma\subseteq X\times A$ be a relation and $\{p(x),\rho_x\}$ be an ensemble of states and their a priori probabilities. Then the tuple $G=(X,A,\sigma,\{p(x),\rho_x\})$ is called a \emph{quantum retrieval game (QRG)}. For a given $x\in X$, an answer $a\in A$ is correct if $(x,a)\in\sigma$.
\end{defn}
The basic setting of a QRG is illustrated in Figure \ref{Fig:diagram}. Notice that minimum-error state discrimination, i.e. the task of correctly identifying the value of $x$ from a measurement of $\rho_x$ with the smallest probability of error, corresponds to the particular case where $\sigma=\{(a,x) : x=a\}$. Therefore, QRGs can be seen as a generalization of state discrimination, but in the case of QRGs, the goal is to determine some information related to $x$, though not necessarily $x$ itself. 

In the most general case, Bob's strategy is defined as a POVM $\{M_a\}_{a\in A}$ with $\sum_a M_a=\id$, $M_a\geq 0$, and where each element $M_a$ is uniquely identified with an answer $a$. The problem of finding the maximum probability of giving a correct answer for a given QRG can be cast as a semi-definite program (SDP) given by
\begin{align*}
\underset{{\{M_a\}}}{\text{maximize  }} &\sum_{(x,a)\in\sigma}p(x)\tr(M_a \rho_x) \\
\text{subject to  } &\sum_a M_a=\id\\
& M_a\geq 0.
\end{align*}
The solution to this SDP, i.e. the maximum probability of giving a correct answer to the QRG, is called the \emph{physical value} $PV(G)$ of the game \cite{gavinsky2012quantum}. Although the physical value of a QRG can always be calculated in principle, it is often preferable to work with a related quantity known as the \emph{selective value} $SV(G)$. It arises by relaxing the condition on the POVM elements to $\sum_a M_a\leq\id$ instead of $\sum_a M_a=\id$, which can intuitively be understood as allowing post-selection in the measurement. Formally, the selective value is defined as the solution of the optimization problem
\begin{align*}
\underset{{\{M_a\}}}{\text{maximize  }} &\frac{\sum_{(x,a)\in\sigma}p(x)\tr(M_a \rho_x)}{\sum_{(x,a)}p(x)\tr(M_a \rho_x)} \\
\text{subject to  } &\sum_a M_a\leq\id\\
& M_a\geq 0.
\end{align*}
Notice that $SV(G)\geq PV(G)$, so an upper bound on the selective value gives also an upper bound on the physical value of the game \cite{pastawski2012unforgeable}. The following is a simple expression for directly calculating the selective value of a QRG.
\begin{theorem}\label{SV_theorem}
\cite{pastawski2012unforgeable} Let $G=(X,A,\sigma,\{p(x),\rho_x\})$ be a QRG. Define 
\beq 
\rho:=\sum_x p(x)\rho_x\nonumber
\eeq 
and
\beq
O_a:=\sum_{x:(x,a)\in\sigma}p(x)\rho^{-\frac{1}{2}}\rho_x\rho^{-\frac{1}{2}}\nonumber.
\eeq
Then $SV(G)=\max_a ||O_a||_{\infty}$.
\end{theorem}
In general, for a particular choice of the sets $X$ and $A$, there are many possible relations $\sigma\in X\times A$. Thus, we can imagine a situation where Bob can choose among many different relations, but is limited in his ability to give a correct answer to all of them. This leads to a natural generalization of QRGs.
\begin{defn}
Let $X,A\subseteq \mathbb{N}$ be respectively the set of inputs and answers, and let $\{p(x),\rho_x\}$ be an ensemble of states. Let also $\sigma_1,\sigma_2,\ldots,\sigma_k\subseteq X\times A$ be different relations and define the QRGs $G_1,G_2,\ldots,G_k$ as before according to these relations. Finally, let $\sigma\subseteq X\times A^k$ be such that $(x,a_1,a_2,\ldots,a_k)\in\sigma$ if and only if $(x,a_i)\in\sigma_i$ for all $i=1,2,\ldots,k$. Then the tuple $G=(X,A^k,\sigma,\{p(x),\rho_x\})$ is called a $\binom{1}{k}$QRG. Additionally, $G$ is called a $(p,\epsilon)-\binom{1}{k}$QRG if it holds that
\begin{enumerate}
\item{$PV(G_i)\geq p$ for all $i=1,2,\ldots,k$}
\item{$PV(G)\leq \epsilon$.}\\
\end{enumerate}
We refer to $p$ as the winning probability and to $\epsilon$ as the cheating probability.
\end{defn}
We also refer to $\binom{1}{k}$QRGs as (1-out-of-k) QRGs. Notice that in (1-out-of-k) QRGs it is possible for Bob to provide a correct answer for any relation with high probability, but it is hard for him to give correct answers to many relations simultaneously. This property makes them useful building-blocks for cryptography.
\subsection{Hidden matching QRGs}
\begin{figure}
 \begin{tikzpicture}[scale=0.35]
  %the multiplication with floats is not possible. Thus I split the loop in two.
  \foreach \number in {1,...,4}{
      % Computer angle:
        \mycount=\number
        \advance\mycount by -4
  \multiply\mycount by -270
        \advance\mycount by 0
      \node[draw,circle,inner sep=0.04cm] (\number) at (\the\mycount:2.3cm) {\number};
    }
\path (1) edge[-,LimeGreen,ultra thick] (2); 
\path (3) edge[-,LimeGreen,ultra thick]  (4);
\begin{scope}[shift={(7.65,0)}]
  \foreach \number in {1,...,4}{
      % Computer angle:
        \mycount=\number
        \advance\mycount by -4
  \multiply\mycount by -270
        \advance\mycount by 0
      \node[draw,circle,inner sep=0.04cm] (\number) at (\the\mycount:2.3cm) {\number};
    }
\path (1) edge[-,NavyBlue,ultra thick] (3); 
\path (2) edge[-,NavyBlue,ultra thick]  (4);
\end{scope}
\begin{scope}[shift={(16.35,0)}]
	\foreach \number in {1,...,4}{
      % Computer angle:
        \mycount=\number
        \advance\mycount by -4
  \multiply\mycount by -270
        \advance\mycount by 0
      \node[draw,circle,inner sep=0.04cm] (\number) at (\the\mycount:2.3cm) {\number};
    }
\path (1) edge[-,BrickRed,ultra thick] (4); 
\path (3) edge[-,BrickRed,ultra thick]  (2);
\end{scope}
\end{tikzpicture}
\caption{The three possible matchings graphs for the case $n=4$: $\{(1,2),(3,4)\}$ (Green), $\{(1,3),(2,4)\}$ (Blue), and $\{(1,4),(2,3)\}$ (Red).}\label{Fig: Matching graphs}
\end{figure}
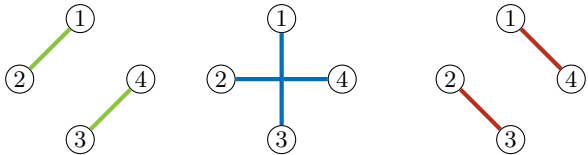

We now focus on a particular class of QRGs which we call \textit{hidden matching} QRGs. A perfect matching $M$ on the set $[n]:=\{1,2,\ldots,n\}$, with $n$ an even number, is a partitioning of $[n]$ into $n/2$ disjoint pairs of numbers. For instance, the three possible matchings for the case $n=4$ are $\{(1,2),(3,4)\}$, $\{(1,3),(2,4)\}$, and $\{(1,4),(2,3)\}$. A matching $M$ on $[n]$ can be represented as a graph with $n/2$ edges, where no two edges share a node in common. We call this the \emph{matching graph} of $n$. This is illustrated in Fig. \ref{Fig: Matching graphs}. 

In hidden matching QRGs, the set of possible inputs is the set of $n$-bit strings, i.e. $X=\{0,1\}^n$, with $n$ an even number. Alice encodes her inputs into the pure states
\beq
\ket{x}=\frac{1}{\sqrt{n}}\sum_{i=1}^n(-1)^{x_i}\ket{i},
\eeq
where $x_i$ is the $i$-th bit of the string $x$ and she prepares each state with probability $\frac{1}{2^n}$.

The set $A$ of possible answers is the set of tuples $(i,j,b)$, where $i,j\in \{1,2,\ldots,n\}$ and $b\in\{0,1\}$. Finally, the relation $\sigma$ is defined as $\sigma=\{(x,i,j,b): x_i\oplus x_j=b\textrm{ and } (i,j)\in M\}$, where $M$ is a perfect matching of $[n]$. In other words, Bob wins the game if he can correctly determine the parity of two bits of $x$ that are joined by an edge in the matching $M$.
The hidden matching problem was first introduced in Ref. \cite{HM-Bar-Yossef} in the context of communication complexity. In that work, it was shown that for any matching there exists a measurement by Bob that can provide a correct answer with certainty. To do so, he measures the states $\ket{x}$ in the basis 
\beq
\{\tfrac{1}{\sqrt{2}}(\ket{i}\pm\ket{j})\},
\eeq
with $(i,j)\in M$. Bob can always provide a correct value because the outcome $\tfrac{1}{\sqrt{2}}(\ket{i}+\ket{j})$ only occurs if $x_i\oplus x_j=0$ and similarly, $\tfrac{1}{\sqrt{2}}(\ket{i}-\ket{j})$ only occurs if $x_i\oplus x_j=1$.

On the other hand, it was shown in Ref. \cite{gavinsky2012quantum} that for the case $n=4$, it is not possible to give a correct answer with certainty for \emph{two} different matchings simultaneously. More precisely, it was shown that there exists a $(1,\frac{3}{4})-\binom{1}{2}$ hidden matching QRG. Interestingly, another $(1,\frac{3}{4})-\binom{1}{2}$ QRG was proposed in Ref. \cite{pastawski2012unforgeable} where Alice encodes her input into sequences of BB84 states.

Moreover, it was shown in Ref. \cite{georgiou2015new} that if the winning and cheating probabilities satisfy the condition
\beq
p>\frac{1+\epsilon}{2},
\eeq
then a $(p,\epsilon)-\binom{1}{2}$ QRG could be repeated $n$ times in order to construct a new $(p',\epsilon')-\binom{1}{2}$ QRG $G'$ with winning probability $p'$ exponentially close to 1 and cheating probability $\epsilon'$ exponentially close to 0. Similarly, it is possible to build (1-out-of-k) QRGs from (1-out-of-2) QRGs by simply combining the smaller QRGs to form a larger one \cite{georgiou2015new}. 

However, the problem with this approach is that in a practical setting, due to experimental imperfections, the winning probability may decrease compared to the ideal case to the point that the QRG is no longer suitable in a cryptographic context. Moreover, the winning probability of a (1-out-of-k) QRG constructed from (1-out-of-2) QRGs will generally decrease with $k$, making this approach less appealing for larger values of $k$. Therefore, it is important to consider the \emph{direct} construction of (1-out-of-k) QRGs for which the winning probability remains high while the cheating probability decreases significantly in $k$. In the following, we generalize the result of Ref. \cite{gavinsky2012quantum} to the case of (1-out-of-k) hidden matching QRGs and show that the probability of correctly answering all relations decreases exponentially in the number of relations $k$, while the winning probability remains unchanged.

In order for these properties to hold, the matchings (and therefore the relations) that we choose to define the QRG must satisfy certain conditions. Consider again the case $n=4$ and suppose that we want to build a (1-out-of-3) QRG from the three possible matchings. From Ref. \cite{gavinsky2012quantum} we know that Bob can give a correct answer for two of the relations with probability at most $3/4$. Would the probability of answering all three relations be even lower than this?

It turns out that this is not the case, since knowing a correct answer to any two relations allows Bob to give a correct answer for the third one with certainty. To see this, suppose for instance that Bob knows that $(x,1,2,0)\in\sigma_1$ and $(x,2,4,1)\in\sigma_2$, i.e. he knows that $x_1\oplus x_2=0$ and $x_2\oplus x_4=1$. From this he can compute $x_1\oplus x_2\oplus x_2\oplus x_4=x_1\oplus x_4=1$, which allows him to give a correct answer $(x,1,4,1)$ for the remaining relation. It is easy to see that this is true for any two different relations and therefore adding a third relation in this case does not decrease Bob's chances to answer all relations correctly.

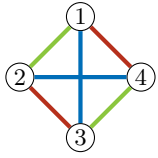
\begin{figure}
 \begin{tikzpicture}[scale=0.35]
  %the multiplication with floats is not possible. Thus I split the loop in two.
  \foreach \number in {1,...,4}{
      % Computer angle:
        \mycount=\number
        \advance\mycount by -4
  \multiply\mycount by -270
        \advance\mycount by 0
      \node[draw,circle,inner sep=0.04cm] (\number) at (\the\mycount:2.3cm) {\number};
    }
\path (1) edge[-,LimeGreen,ultra thick] (2); 
\path (3) edge[-,LimeGreen,ultra thick]  (4);
\path (1) edge[-,BrickRed,ultra thick] (4); 
\path (3) edge[-,BrickRed,ultra thick]  (2);
\path (1) edge[-,NavyBlue,ultra thick] (3); 
\path (2) edge[-,NavyBlue,ultra thick]  (4);
\end{tikzpicture}
\caption{The joint graph for the case $n=4$. The edges are coloured differently to emphasize which matching graph they belonged to originally. This joint graph has many 3-cycles where each edge in the cycle originates from a different matching. For example, the edges $(1,2),(2,3)$ and $(3,1)$ form such a 3-cycle. From Observation \ref{Obs: Cycles}, this means that if Bob can answer two relations correctly, he can answer the third one with certainty.}\label{Fig: joint graph}
\end{figure}

We can understand this more easily by looking at the matching graphs. Suppose we have $k$ different matchings $M_1,M_2,\ldots, M_k$. As illustrated in Fig. \ref{Fig: joint graph}, we can ``join" the matching graphs by constructing a new graph that includes all edges from the original graphs. We call this the \emph{joint} graph. We can make a connection between the existence of closed cycles in the joint graph and the possibility of constructing a (1-out-of-k) QRG from these matchings.
\begin{obs}\label{Obs: Cycles}
Let $M_1,M_2,\ldots, M_k$ be different matchings of the set $[n]:=\{1,2,\ldots,n\}$ and let $\sigma_1,\sigma_2,\ldots,\sigma_k$ be the relations obtained from them in a hidden matching QRG. Let also $\Gamma_k$ be the joint graph obtained from the corresponding matching graphs. If there exists a $k$-cycle in the joint graph such that each edge in the cycle belonged to a different matching graph, then there exists a set of correct answers for $k-1$ relations that allows Bob to provide a correct answer for the $k$-th relation with certainty.
\end{obs} 
\textit{Proof:} A $k$-cycle is a closed loop formed by traversing nodes connected by $k$ edges, starting and ending in the same node. Suppose that there exists a $k$-cycle in the joint graph $\Gamma_k$ formed by edges that originally belonged to different matching graphs. Furthermore, without loss of generality, assume that the cycle begins and ends in node $i$ and that the sequence of edges is $(i,j_1),(j_1,j_2),\ldots, (j_{k-2},j_{k-1}),(j_{k-1},i)$. 
Now suppose that Bob knows a correct answer to the $k-1$ relations $\sigma_1,\sigma_2,\ldots,\sigma_{k-1}$ formed from matchings that include the edges $(i,j_1),(j_1,j_2),\ldots, (j_{k-2},j_{k-1})$. This means that Bob also knows the values $x_i\oplus x_{j_1}, x_{j_1}\oplus x_{j_2},\ldots x_{j_{k-2}}\oplus x_{j_{k-1}}$ and by taking the modulo 2 sum of these, he obtains the value of $x_i\oplus x_{j_{k-1}}$. This allows him to provide a correct answer to the remaining relation, since $(i, j_{k-1},x_i\oplus x_{j_{k-1}})\in\sigma_k$. \qed

The above observation tells us that to construct (1-out-of-k) QRGs, we must choose the matchings carefully in order avoid cycles in the joint graph. This leads us to the following definition.
\begin{defn}
A set of $k$ matchings $M_1,M_2,\ldots, M_k$ are called \emph{independent} if their corresponding joint graph $\Gamma_k$ does not have any $k$-cycles whose edges correspond to different matchings.
\end{defn}
We now show that the probability of giving correct answers to all relations in a (1-out-of-k) hidden matching QRG decays exponentially in $k$ provided that the matchings used are independent.
\begin{theorem}
Let $G$ be a (1-out-of-k) hidden matching QRG, where the $k$ different matchings are independent. Then the winning probability $p$ and the cheating probability $\epsilon_k$ satisfy
\begin{align}
p &=1\\
\epsilon_k &\leq\frac{k+1}{2^k}.
\end{align}
\end{theorem} 
\textit{Proof:} Recall that the states used in a hidden matching QRG are 
\beq
\ket{x}=\frac{1}{\sqrt{n}}\sum_{i=1}^n(-1)^{x_i}\ket{i},
\eeq
with corresponding density matrices $\rho_x=\ketbra{x}{x}$ and each state is prepared with probability $\frac{1}{2^n}$. As discussed before, for hidden matching QRGs, it is possible to give a correct answer for any given relation with certainty, so we have that $p=1$.

We want to use the result of theorem \ref{SV_theorem} to calculate the selective value of the game. It holds that
\beq\label{rho}
\rho:=\frac{1}{2^n}\sum_x \rho_x=\frac{1}{n}\id
\eeq
and
\begin{align}
\rho^{-\frac{1}{2}}\rho_x\rho^{-\frac{1}{2}}&=n\rho_x\nonumber\\
&=\sum_{i,j}(-1)^{x_i\oplus x_j}\ketbra{i}{j}:=\Pi_x.
\end{align}
The operators $O_a$ take the form
\beq
O_a=\frac{1}{2^n}\sum_{x:(x,a)\in\sigma}\sum_{i=1}^n\sum_{j=1}^n(-1)^{x_i\oplus x_j}\ketbra{i}{j}
\eeq
and
\beq
\left[O_a\right]_{ij}=\frac{1}{2^n}\sum_{x:(x,a)\in\sigma}(-1)^{x_i\oplus x_j}
\eeq
where $\left[O_a\right]_{ij}$ is the entry of the $i$-th row and $j$-th column of the matrix representation of $O_a$.
In a (1-out-of-k) hidden matching QRG, $(x,a)\in\sigma$ if and only if $(x,a_l)\in \sigma_l$ for all $l=1,2,\ldots,k$ and $(x,a_l)\in \sigma_l$ if and only if $x_i\oplus x_j=b$ and $(i,j)\in M_l$, where $M_l$ is the matching defining the relation $\sigma_l$. Each relation $\sigma_l$ induces a constraint on the possible values of $x \in \{0,1\}^n$. Since the matchings are independent, each individual relation $\sigma_l$ cuts down by half the number of strings $x$ that satisfy $(x,a)\in\sigma$. Therefore, for every $a$, the total number of strings satisfying $(x,a)\in\sigma$ is equal to $2^{n-k}$.

From this we can easily calculate the diagonal elements of $O_a$ to be $\left[O_a\right]_{ii}=2^{-k}$. Let $a=(i_1,j_1,b_1),(i_2,j_2,b_2),\ldots,(i_k,j_k,b_k)$ and define the set $S:=\{(i_1,j_1),(i_2,j_2),\ldots,(i_k,j_k)\}$. Then it holds that $\left[O_a\right]_{i_lj_l}=2^{-k}(-1)^{b_l}$ for all $l=1,2,\ldots,k$ since we must have that $x_{i_l}\oplus x_{j_l}=b_l$. From an identical argument as in Observation \ref{Obs: Cycles}, if the edges in $S$ can form a path that connects two nodes $(i',j')$, this fixes the value of $x_{i'}\oplus x_{j'}$. Let $P$ be the set of all such edges $(i',j')$. Finally, for any other edge $(i,j)$ that has no constraints, $x_i\oplus x_j=0$ for half of the values of $x$ and $x_i\oplus x_j=1$ for the other half. Thus, we can conclude that
\beq\label{EQ: O_a}
\left[O_a\right]_{ij}=\left\{ \begin{array}{cc}
2^{-k} & \textrm{ if } i=j.  \\
\pm 2^{-k} & \textrm{ if } (i,j)\in S\cup P\\
0 & \textrm{ otherwise.}
\end{array} \right. 
\eeq
Each answer $a=(i_1,j_1,b_1),\ldots,(i_k,j_k,b_k)$ defines a graph of $n$ nodes with edges $(i_1,j_1),\ldots,(i_k,j_k)$. This graph can be divided into subgraphs, where each subgraph is defined by all edges that form a connected path. By construction, these subgraphs are disconnected from each other. Thus, for any two nodes $i$ and $j$ that belong to different subgraphs, Eq. \eqref{EQ: O_a} implies that $\left[O_a\right]_{ij}=0$, which means that $O_a$ must be block-diagonal in the subgraphs. Moreover, since $O_a$ is positive, each block must also be positive. The largest possible block has dimension $k+1$ and from Eq. \eqref{EQ: O_a}, its trace is equal to $(k+1)2^{-k}$, so its largest possible eigenvalue is $(k+1)2^{-k}$. This is also the largest possible eigenvalue of $O_a$, so we have that
\beq
SV=\max_a ||O_a||_{\infty}=\frac{k+1}{2^k}.
\eeq
Since the physical value of the game is upper bounded by the selective value, we conclude that
\beq
\epsilon\leq \frac{k+1}{2^k}
\eeq 
as desired. \qed

From this theorem, we know that we can construct (1-out-of-k) QRGs as long as we can create hidden matching QRGs that have $k$ independent matchings. We now provide a method of constructing sets of independent matchings.

Suppose we start with a set of $k$ independent matchings on a graph with $n$ nodes. We can create a new set of $k+1$ independent matchings on a graph with $2n$ nodes in the following way: Take the joint graph of the $k$ independent matchings and generate an isomorphic graph by simply re-labelling all the nodes. For each node $i$ in the first graph, make a new edge by connecting it to a unique node $j$ in the second graph. This new collection of edges forms a new matching, so we have $k+1$ matchings in total on a larger graph with $2n$ nodes.

To see that these matchings must be independent, notice that the new matching is built by connecting nodes from two separate graphs, each of which is the joint matching graph of $k$ independent matchings. By definition of independent matchings, there are no closed cycles in either of these two graphs, so we cannot create a closed cycle by adding an edge between them. Therefore, the $k+1$ matchings must be independent.

Using this method, we can provide explicit examples of $k$ independent matchings. If we start with the case $n=4$ and the two independent matchings $\{(1,2),(3,4)\}$ and $\{(1,3),(2,4)\}$, we can apply the canonical construction to obtain $k$ independent matchings with $n=2^k$. This is illustrated in Fig. \ref{Fig: Canonical}. Every matching in this construction has the property that the distance between any two nodes connected by an edge is constant. More precisely, the edges in each matching are of the form $(i,i+2^{j-1})$ with $j=1,2,\ldots,k.$ As we discuss in section \ref{Implementation}, this feature greatly helps in their implementation.

Alternatively, we can start with a set of 3 independent matchings for $n=6$, namely $\{(1,2),(3,4),(5,6)\}$, $\{(1,6),(2,3),(4,5)\}$ and $\{(1,4),(2,5),(3,6)\}$. As seen in Fig. \ref{Fig: 6,12}, using the method of the canonical construction we can obtain $k$ independent matchings for $n=3\times 2^{k-2}$ and $k\geq 3$.
\begin{center}
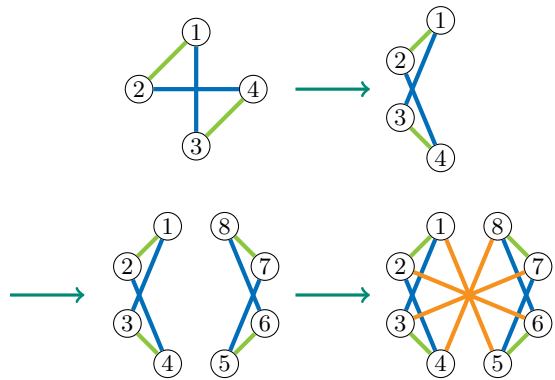
\begin{figure}
%----FIRST GRAPH -------
\begin{tikzpicture}[scale=0.33]
\begin{scope}[shift={(7,0)}]
  \foreach \number in {1,...,4}{
      % Computer angle:
        \mycount=\number
        \advance\mycount by -4
  \multiply\mycount by -270
        \advance\mycount by 0
      \node[draw,circle,inner sep=0.038cm] (\number) at (\the\mycount:2.3cm) {\number};
    }
 
\path (1) edge[-,LimeGreen,ultra thick] (2); 
\path (3) edge[-,LimeGreen,ultra thick]  (4);

\path (2) edge[-,NavyBlue,ultra thick] (4); 
\path (3) edge[-,NavyBlue,ultra thick]  (1);
\end{scope}

%------ ARROW -----------
\draw [->,very thick,PineGreen] (11,0) -- (14,0);

%----- UNFOLDED GRAPH ------------
\begin{scope}[shift={(18,0)},rotate=22.5]

\foreach \number in {1,...,4}{
      % Computer angle:
        \mycount=\number
        \advance\mycount by 1
  \multiply\mycount by 45
        \advance\mycount by 0
      \node[draw,circle,inner sep=0.038cm] (\number) at (\the\mycount:3cm) {\number};
    }

\path (2) edge[-,LimeGreen,ultra thick] (1); 
\path (4) edge[-,LimeGreen,ultra thick]  (3);
\path (1) edge[-,NavyBlue,ultra thick] (3); 
\path (2) edge[-,NavyBlue,ultra thick] (4); 

\end{scope}

%---- LOW ARROW #1 ----------
\draw [->,very thick, PineGreen] (-0.5,-8.3) -- (2.5,-8.3);

% -------- TWO UNFOLDED GRAPHS -------------

\begin{scope}[shift={(7,-8.3)},rotate=22.5]

\foreach \number in {1,...,8}{
      % Computer angle:
        \mycount=\number
        \advance\mycount by 1
  \multiply\mycount by 45
        \advance\mycount by 0
      \node[draw,circle,inner sep=0.038cm] (\number) at (\the\mycount:3cm) {\number};
    }

\path (5) edge[-,LimeGreen,ultra thick] (6); 
\path (7) edge[-,LimeGreen,ultra thick] (8); 

\path (8) edge[-,NavyBlue,ultra thick] (6); 
\path (7) edge[-,NavyBlue,ultra thick] (5);

\path (2) edge[-,LimeGreen,ultra thick] (1); 
\path (4) edge[-,LimeGreen,ultra thick]  (3);
\path (1) edge[-,NavyBlue,ultra thick] (3); 
\path (2) edge[-,NavyBlue,ultra thick] (4); 

\end{scope}

%---- LOW ARROW #2 ----------
\draw [->,very thick, PineGreen] (11,-8.3) -- (14,-8.3);

% -------- JOINT GRAPH -------------

\begin{scope}[shift={(18,-8.3)},rotate=22.5]
\foreach \number in {1,...,8}{
      % Computer angle:
        \mycount=\number
        \advance\mycount by 1
  \multiply\mycount by 45
        \advance\mycount by 0
      \node[draw,circle,inner sep=0.038cm] (\number) at (\the\mycount:3cm) {\number};
    }

\path (5) edge[-,LimeGreen,ultra thick] (6); 
\path (7) edge[-,LimeGreen,ultra thick] (8); 

\path (8) edge[-,NavyBlue,ultra thick] (6); 
\path (7) edge[-,NavyBlue,ultra thick] (5);

\path (2) edge[-,LimeGreen,ultra thick] (1); 
\path (4) edge[-,LimeGreen,ultra thick]  (3);
\path (1) edge[-,NavyBlue,ultra thick] (3); 
\path (2) edge[-,NavyBlue,ultra thick] (4); 

\path (5) edge[-,BurntOrange,ultra thick] (1); 
\path (7) edge[-,BurntOrange,ultra thick]  (3);
\path (4) edge[-,BurntOrange,ultra thick] (8); 
\path (2) edge[-,BurntOrange,ultra thick] (6); 
\end{scope}
\end{tikzpicture}
\caption{Canonical construction of independent matchings. We begin with the joint graph of two independent matchings for the case $n=4$. The joint graph is then combined with an isomorphic version of it to form a graph of 8 nodes. For $i=1,2,3,4$, node $i$ in the new graph is joined with the node $i+4$ to form a new graph with three independent matchings. This process can be repeated indefinitely to produce any $k$ independent matchings with $n=2^k$.  }\label{Fig: Canonical}
\end{figure}
\end{center}
In summary, we have outlined a method of constructing (1-out-of-k) QRGs from the hidden matching problem and shown that their cheating probability decreases exponentially with $k$. In order to achieve this, the matchings used must be independent and we provide a constructive method of generating independent matchings. In the next section, we study how these QRGs can be implemented in practice.
\begin{center}
\begin{figure}[t]
\begin{tikzpicture}[scale=0.45]
  %the multiplication with floats is not possible. Thus I split the loop in two.
  \foreach \number in {1,...,6}{
      % Computer angle:
        \mycount=\number
        \advance\mycount by 1
  \multiply\mycount by 60
        \advance\mycount by 0
      \node[draw,circle,inner sep=0.065cm] (\number) at (\the\mycount:3.0cm) {\number};
    }
 
\path (1) edge[-,BrickRed,ultra thick] (2); 
\path (3) edge[-,BrickRed,ultra thick]  (4);
\path (5) edge[-,BrickRed,ultra thick] (6); 
\path (1) edge[-,NavyBlue,ultra thick]  (6);
\path (5) edge[-,NavyBlue,ultra thick] (4); 
\path (3) edge[-,NavyBlue,ultra thick]  (2);
\path (1) edge[-,BurntOrange,ultra thick] (4); 
\path (2) edge[-,BurntOrange,ultra thick]  (5);
\path (3) edge[-,BurntOrange,ultra thick] (6);

\begin{scope}[shift={(10,0)},rotate=45]

\foreach \number in {1,...,9}{
      % Computer angle:
        \mycount=\number
        \advance\mycount by 1
  \multiply\mycount by -330
        \advance\mycount by 0
      \node[draw,circle,inner sep=0.10cm] (\number) at (\the\mycount:4.1cm) {\number};
    }
  \foreach \number in {10,...,12}{
      % Computer angle:
        \mycount=\number
        \advance\mycount by 1
  \multiply\mycount by -330
        \advance\mycount by 0
      \node[draw,circle,inner sep=0.065cm] (\number) at (\the\mycount:4.1cm) {\number};
    }

\path (1) edge[-,BrickRed,ultra thick] (2); 
\path (3) edge[-,BrickRed,ultra thick]  (4);
\path (5) edge[-,BrickRed,ultra thick] (6); 
\path (7) edge[-,BrickRed,ultra thick] (8); 
\path (9) edge[-,BrickRed,ultra thick]  (10);
\path (11) edge[-,BrickRed,ultra thick] (12); 

\path (2) edge[-,NavyBlue,ultra thick] (3); 
\path (4) edge[-,NavyBlue,ultra thick]  (5);
\path (6) edge[-,NavyBlue,ultra thick] (7); 
\path (8) edge[-,NavyBlue,ultra thick] (9); 
\path (10) edge[-,NavyBlue,ultra thick]  (11);
\path (12) edge[-,NavyBlue,ultra thick] (1); 

\path (1) edge[-,BurntOrange,ultra thick]  (4);
\path (2) edge[-,BurntOrange,ultra thick]  (5); 
\path (3) edge[-,BurntOrange,ultra thick]  (6);
\path (7) edge[-,BurntOrange,ultra thick]  (10); 
\path (8) edge[-,BurntOrange,ultra thick]  (11);
\path (9) edge[-,BurntOrange,ultra thick]  (12);

\path (1) edge[-,LimeGreen,ultra thick]  (7); 
\path (2) edge[-,LimeGreen,ultra thick]  (8);
\path (3) edge[-,LimeGreen,ultra thick]  (9);
\path (4) edge[-,LimeGreen,ultra thick]  (10); 
\path (5) edge[-,LimeGreen,ultra thick]  (11);
\path (6) edge[-,LimeGreen,ultra thick]  (12);
\end{scope}
\end{tikzpicture}
\caption{Joint graph of 3 independent matchings for the case $n=6$ and joint graph of 4 independent matchings for the case $n=12$. The procedure of the canonical construction can be followed to obtain any $k$ independent matchings with $n=3\times 2^{k-2}$.}\label{Fig: 6,12}
\end{figure}
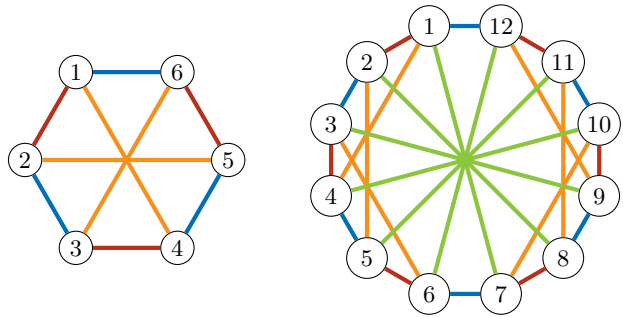
\end{center}
\section{Coherent-state QRGs}\label{Implementation}
An implementation of (1-out-of-k) hidden matching QRGs faces two main challenges: the high dimensionality of the quantum states and the complex measurement by Bob. For the constructions we have outlined in the previous section, the dimension of the states scales exponentially with $k$, namely as $2^k$ for the canonical construction. On the other hand, Bob must be able to perform $k$ different measurements corresponding to each possible matching, each of which involves projections onto coherent superpositions of pairs of basis states. Given these obstacles, we propose new QRGs based on coherent states of light. These games use a signal structure developed in Ref. \cite{arrazolaqfp} with a mapping of the qubit signals into the coherent state signals following the procedures of Ref. \cite{arrazola2014QC}. 

In this construction, the states
\beq
\ket{x}=\frac{1}{\sqrt{n}}\sum_{i=1}^n(-1)^{x_i}\ket{i}\nonumber
\eeq
are mapped to a sequence of coherent states
\beq
\ket{\alpha, x}=\bigotimes_{i=1}^{n}\left|(-1)^{x_i}\frac{\alpha}{\sqrt{n}}\right\rangle
\eeq
where $|\alpha|^2$ is the total mean photon number, which is a freely chosen parameter. We consider the modes to be separated in time on a single spatial mode, i.e. they are time-bin modes. Notice that the bit values of $x$ are encoded in the phase of the corresponding coherent state and all coherent states have the same amplitude, namely $\frac{\alpha}{\sqrt{n}}$.

The measurement by Bob is carried out by first applying a permutation of the optical modes. The purpose of this is to pair modes together according to the matching, which are later used as input to a balanced beam splitter. More precisely, for every $(i,j)\in M$, Bob pairs modes $i$ and $j$ that later interfere in a balanced beam-splitter. He then measures the output modes with two single-photon threshold detectors, which are labelled $D_0$ and $D_1$.

If the incoming states to the beam splitter are 
\beq
\ket{(-1)^{x_i}\tfrac{\alpha}{\sqrt{n}}}\otimes\ket{(-1)^{x_j}\tfrac{\alpha}{\sqrt{n}}},
\eeq
the output states are
\beq\label{Eq: coherent states}
\ket{\left(1+(-1)^{x_i\oplus x_j}\right)\tfrac{\alpha}{\sqrt{n}}}\otimes\ket{\left(1-(-1)^{x_i\oplus x_j}\right)\tfrac{\alpha}{\sqrt{n}}}.
\eeq
Therefore, for each possible value of $x_i\oplus x_j$, the output states are either a vacuum state (which never produces a click), or a weak coherent state with amplitude $\frac{\alpha}{\sqrt{n}}$, which creates a click with non-zero probability. In the ideal case, only one detector $D_0$ can click if $x_i\oplus x_j=0$ while only the other detector $D_1$ can click if $x_i\oplus x_j=1$. Thus, whenever there is a click, Bob can provide a correct answer to the relation corresponding to his chosen matching. In fact, it is possible to obtain many clicks corresponding to different elements of the matching, and they all correspond to a correct answer. The only issue that can arise is that no clicks occur, in which case Bob gets no information and has to guess the answer at random. The probability that at least one click occurs, i.e. the winning probability, can be calculated from the photon statistics of the states, leading to the expression
\beq\label{EQ: winning prob. }
p=1-\frac{1}{2}e^{-|\alpha|^2}.
\eeq
In general, it is difficult to perform an arbitrary permutation of a large number of optical modes. However, in our case, if we focus on (1-out-of-k) hidden matching QRGs built from the canonical construction, all relevant matchings have the property that the distance between nodes connected by an edge is constant. For example, for a (1-out-of-3) QRG with $n=8$, in the matching $M_{1,3}=\{(1,2),(3,4),(5,6),(7,8)\}$ illustrated in Fig. \ref{Fig: Canonical}, all nodes connected by an edge have distance 1. Similarly, for $M_{2,3}$ they have distance 2 and for $M_{3,3}$ they have distance 4. In general, the distance between all nodes connected by an edge in a matching $M_{j,k}$ is $2^{j-1}$.

\begin{center}
\begin{figure}
\includegraphics[width=\columnwidth]{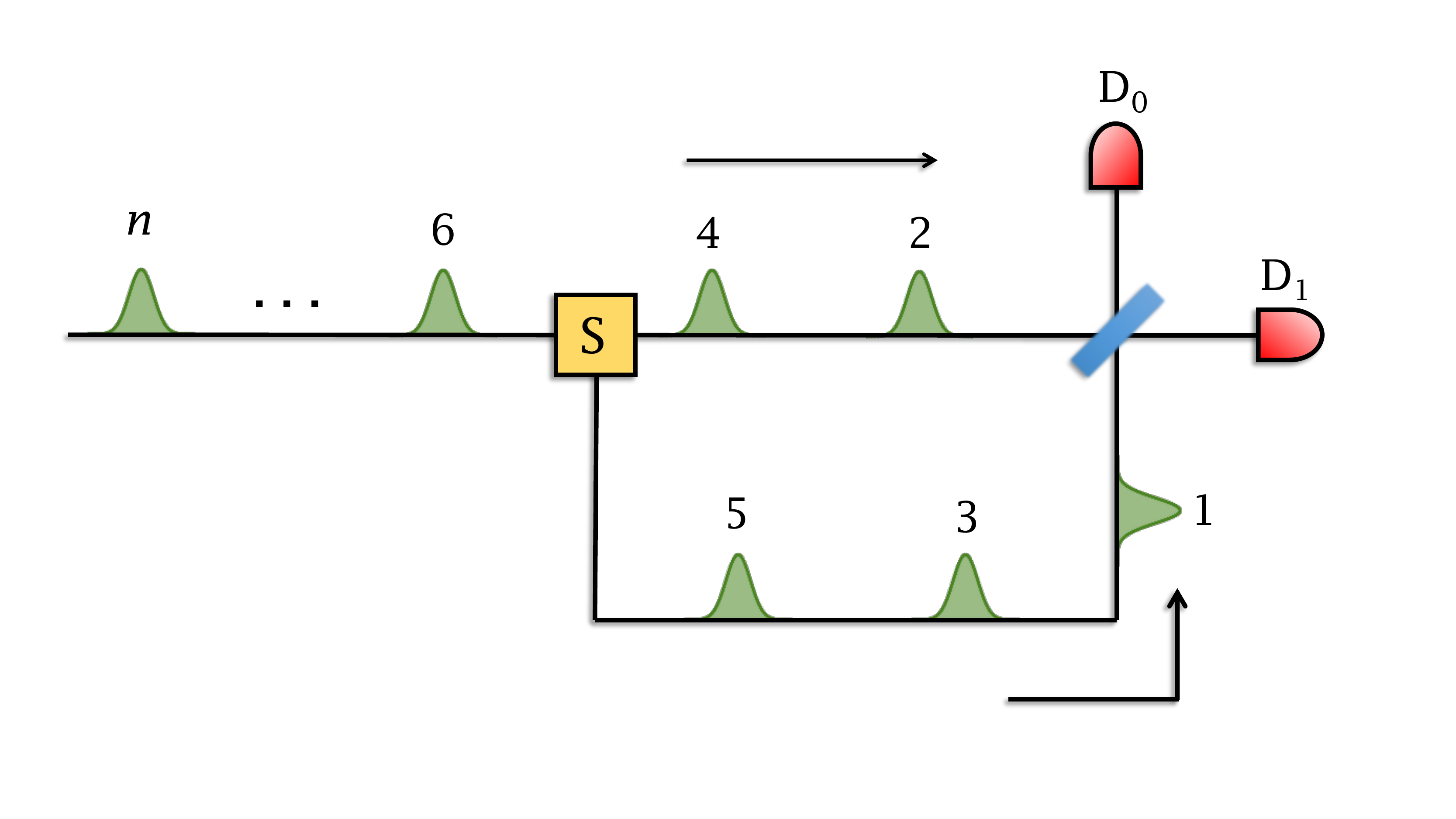}
\caption{Implementation of hidden matching QRGs using coherent states and linear optics. Alice prepares a sequence of $n$ coherent states and sends them to Bob, who uses a switch to choose which path of an interferometer each pulse takes. At the end, the pulses are paired according to the matching and they interfere in pairs in a balanced beam splitter. In this case, we consider the matching $M_{1,k}$, where the pulses that interfere are separated by one time slot. The outputs are measured with single-photon detectors and whenever there is a click, Bob knows the parity of the corresponding bits.}\label{Fig: Implementation}
\end{figure}
\end{center}

From an implementation's point of view, this means that Bob only needs the ability to selectively displace incoming modes by a \textit{fixed} amount of time. He can achieve this by using an active optical switch, which directs incoming pulses into either one of two arms of an interferometer. The path length of the long arm of the interferometer is set to differ from the short one in exactly the amount needed to displace the pulse in time, allowing him to interfere the intended pulses. If Bob wants an answer for a different relation, he only has to change the path length of the long arm of the interferometer. The setup is illustrated in Fig. \ref{Fig: Implementation}.

\begin{center}
\begin{figure}[t!]
\includegraphics[width=\columnwidth]{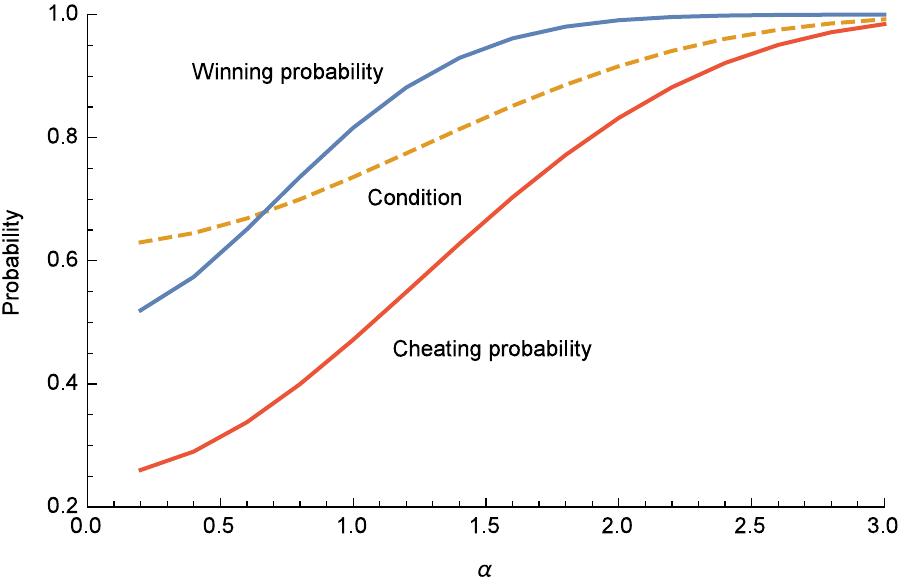}
\caption{Winning and cheating probability for a coherent-state (1-out-of-2) QRG in an ideal implementation. The gap between the winning and cheating probability can be adjusted by choosing the parameter $\alpha$ appropriately. For a wide range of values of $\alpha$, the winning probability $p$ and the cheating probability $\epsilon$ satisfy the condition $p>(1+\epsilon)/2$ required to construct a (1-out-of-2) QRG with exponentially small cheating probability \cite{georgiou2015new}.}\label{Fig: QRG2}
\end{figure}
\end{center}

\begin{center}
\begin{figure}[t!]
\includegraphics[width=\columnwidth]{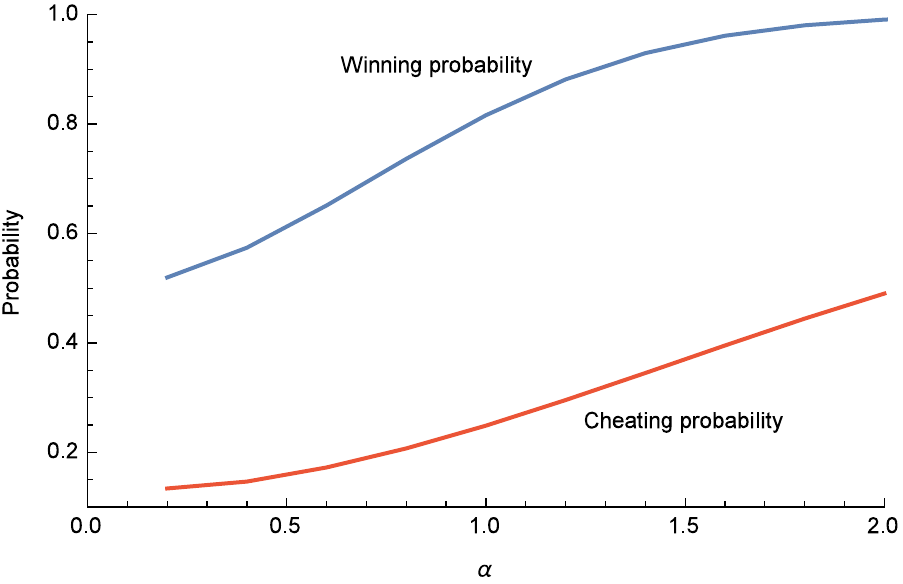}
\caption{Winning and cheating probability for a coherent-state (1-out-of-3) QRG with $n=6$ in an ideal implementation. The gap between the winning and cheating probability can be adjusted by choosing the parameter $\alpha$ appropriately.}\label{Fig: QRG3}
\end{figure}
\end{center}

Although the optical switch in Bob's system is an active optical element that requires fast switching, it can be achieved using Mach-Zehnder switching. Additionally, notice that in this setup, Alice has the ability to prepare states corresponding to very large values of $n$, so the scalability of the system is mostly limited by the interferometer, where there will be a limit to the length of the long arm that can be achieved while maintaining stability. Despite this fact, this configuration can be used to implement (1-out-of-k) QRGs for values of $k$ that exceed what could be feasibly done in a qubit implementation.
It must be noted that coherent-state QRGs are not identical to those studied in section \ref{Section: QRGs}. Nevertheless, as shown in Ref. \cite{arrazola2014QC}, the coherent-state version of quantum communication protocols retain the essential properties of the original ones. We can show this explicitly by directly calculating the semi-definite program (SDP) corresponding to the coherent-state QRGs to obtain a value for the cheating probability. Recall that the cheating probability is the probability that Bob gives a correct answer to \textit{all} relations. Thus, the SDP corresponds to the problem
\begin{align*}
\underset{{\{M_a\}}}{\text{maximize  }} &\sum_{(x,a)\in\sigma}\frac{1}{2^n}\tr(M_a \ketbra{\alpha,x}{\alpha,x}) \\
\text{subject to  } &\sum_a M_a=\id\\
& M_a\geq 0.
\end{align*}
where the states $\ket{\alpha,x}$ are the same as in Eq. \eqref{Eq: coherent states}, $a=\{a_1,a_2,\ldots,a_k\}$ and where $(x,a)\in\sigma$ if and only if $(x,a_i)\in\sigma_i$ for all $i=1,2,\ldots,k$. The value of this SDP can be calculated numerically, but for large values of $k$ it becomes computationally demanding since the dimension of the states and the number of variables of the SDP grow very quickly. For this reason, we have calculated values of the cheating probability for a (1-out-of-2) QRG obtained from the canonical construction and a (1-out-of-3) QRG for the case $n=6$ using the matchings of Fig. \ref{Fig: 6,12}. These are shown respectively in Figs. \ref{Fig: QRG2} and \ref{Fig: QRG3}.

As it can be seen in Figs. \ref{Fig: QRG2} and \ref{Fig: QRG3}, by choosing the value of the parameter $\alpha$ appropriately, it is possible to achieve large gaps between the winning and cheating probability of the QRG. For example, in the case of the (1-out-of-2) QRG, the gap is large enough to satisfy the condition $p>(\epsilon+1)/2$ of Ref. \cite{georgiou2015new}.

As mentioned in section \ref{Section: QRGs}, experimental imperfections may not permit a large enough difference between the winning and error probabilities. In particular, in a cryptographic context, an adversary is limited only by quantum mechanics and can therefore always achieve the cheating probability, whereas the winning probability of an honest player may be limited by experimental imperfections. Thus, understanding the role of imperfections is crucial for practical applications of QRGs.
\subsection{Experimental imperfections}
In the implementation of coherent-state QRGs there are two main sources of imperfections: the combined effect of transmission loss and detector efficiency, as well the limited visibility of the interferometer, to which we associate the parameters $\eta$ and $\nu$ respectively. Dark counts are not significant, since we are dealing with coherent states with parameter $\alpha\sim 1$, while dark count probabilities of less than $10^{-3}$ can be easily obtained.

The effect of loss and limited efficiency is to change the parameter $\alpha$ of the states to $\sqrt{\eta}\alpha$, with  $0\leq\eta\leq 1$. This leads to a reduction in the probability of observing at least one photon, which in turn reduces the winning probability. Additionally, if the visibility of the interferometer is limited, it is possible for the wrong detector to click at each time slot (leading to an incorrect answer) or for both to detectors click, in which case Bob has to guess the answer at random and can also make a mistake. 

In the presence of these imperfections, we can write the output of the beam-splitter at each time slot as the state
\beq
\left|\sqrt{\frac{2\eta\nu}{n}}\alpha\right\rangle\otimes\left|\sqrt{\frac{2\eta(1-\nu)}{n}}\alpha\right\rangle,
\eeq
with $0\leq\nu\leq 1$ and where we assume the first mode corresponds to the correct detector. Hence, large values of $\nu$ imply that the interference works properly and most of the photons go to the correct output mode. Notice that $\nu$ can be related to the usual visibility reported in experiments, which is given by $2\nu-1$. 

The probability that there is a click in the correct detector is 
\beq
p_c=1-\exp\left(-2\eta\nu \frac{|\alpha|^2}{n}\right)
\eeq
and the probability of obtaining a click in the wrong detector is
\beq
p_w=1-\exp\left(-2\eta(1-\nu) \frac{|\alpha|^2}{n}\right).
\eeq
Over the entire run of the experiment, Bob may give an incorrect value if no clicks occur. The probability that this happens is given by
\beq\label{Eq: p0}
p_0=\exp\left(-\eta|\alpha|^2\right).
\eeq
For simplicity, we consider the case where Bob selects at random between all slots where at least one click occurs, even if they were double clicks. When there are double clicks, Bob simply guesses an answer at random. Notice that this gives us a lower bound on the winning probability that would be obtained if Bob discards double clicks at one time slot if single clicks occurred at another time slot.
\begin{center}
\begin{figure}
\includegraphics[width=\columnwidth]{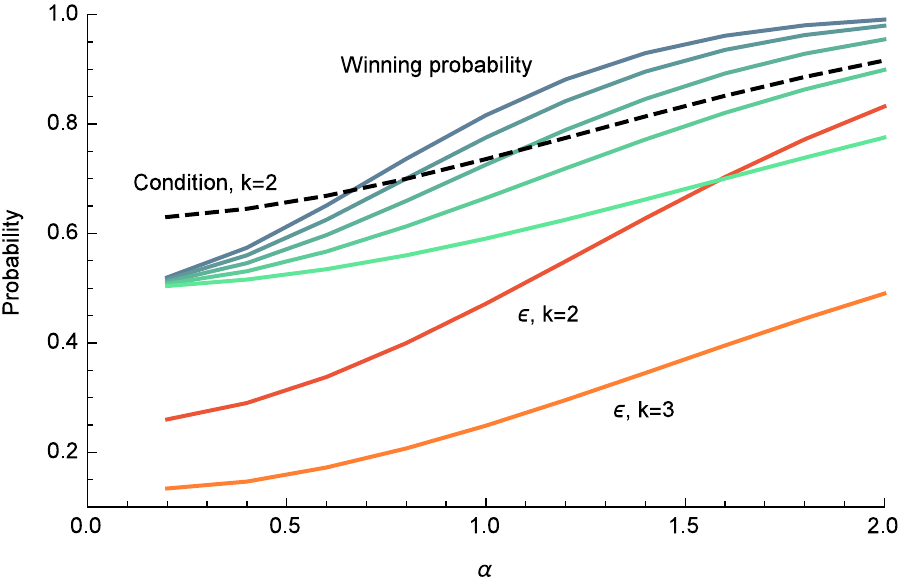}
\caption{Winning probability in the presence of limited efficiency. We consider the case $\nu=1$ and plot the winning probabilities as a function of $\alpha$ for values $\eta=1,0.8,0.6,0.4,0.2$. We also show the cheating probability for both $k=2$ and $k=3$, as well as the condition $p>(1+\epsilon)/2$ for the case $k=2$. The winning probability decreases with lower $\eta$ and for values smaller than $\eta=0.6$ it is not possible to satisfy this condition.}\label{Fig: ExpEta}
\end{figure}
\end{center}
\begin{center}
\begin{figure}
\includegraphics[width=\columnwidth]{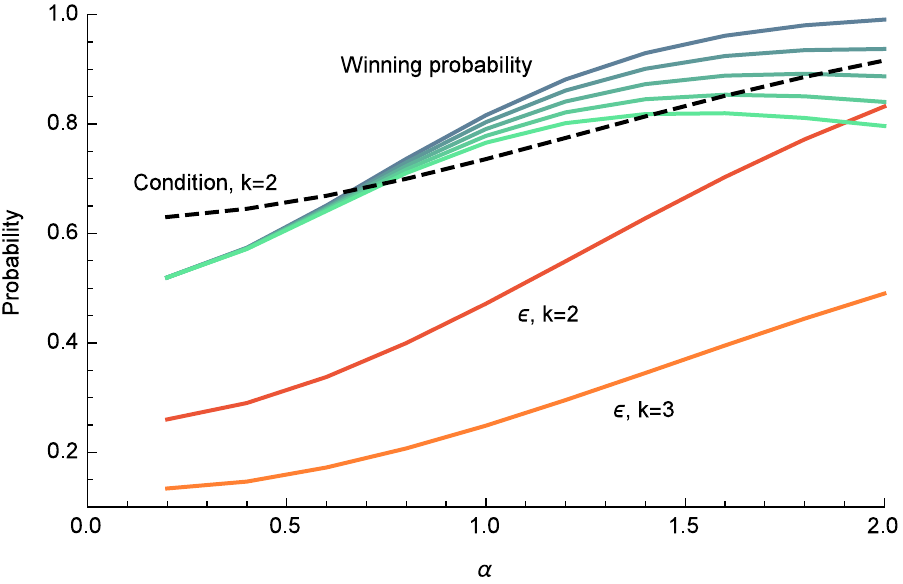}
\caption{Winning probability in the presence of limited visibility. We consider the case $\eta=1$ and plot the winning probabilities as a function of $\alpha$ for values $\nu=1,0.95,0.9,0.85,0.8$. The winning probability decreases with lower $\nu$ but for this range of values of $\nu$ it is always possible to choose an $\alpha$ such that the condition $p>(1+\epsilon)/2$ is satisfied.}\label{Fig: ExpNu}
\end{figure}
\end{center}
\begin{center}
\begin{figure}[t!]
\includegraphics[width=\columnwidth]{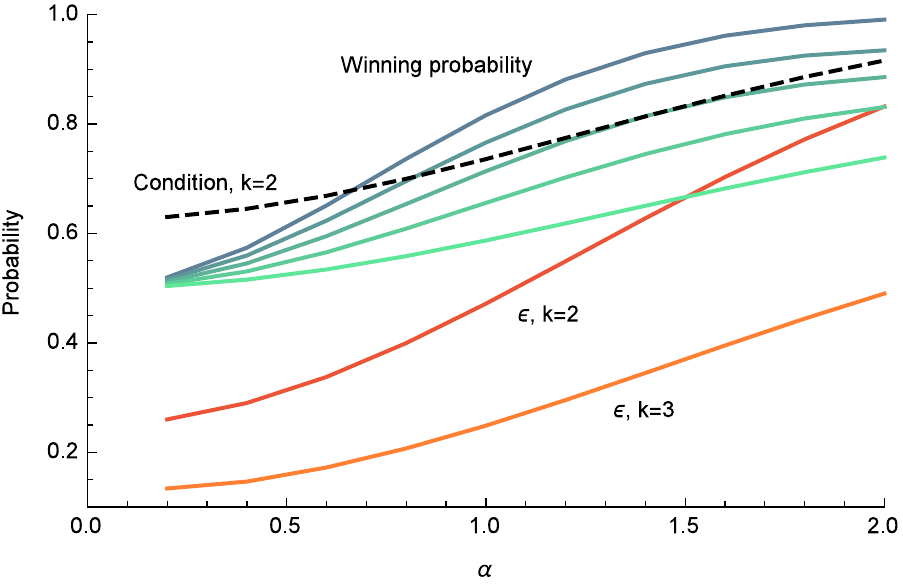}
\caption{Winning probability in the presence of limited visibility and efficiency. We plot the winning probabilities for different values of both $\eta$ and $\nu$, namely $(\nu,\eta)=(1,1)$, (0.95,0.8), (0.9,0.6), (0.85,0.4) and (0.8,0.2). When both experimental imperfections are present it is harder to satisfy the condition $p>(1+\epsilon)/2$, but it is still possible for example for the values $\eta=0.8$ and $\nu=0.95$. However, the gap with respect to the cheating probability for $k=3$ remains large even with significant imperfections.}\label{Fig: ExpEtaNu}
\end{figure}
\end{center}
In this case, the probability of giving a wrong answer when there is at least one click is equal to the probability of giving a wrong answer for a single time slot \textit{given that there was a click in that slot}, since these probabilities are equal for each slot.  Thus, the probability of error given that there is at least one click is 
\beq\label{Eq: p1}
p_1=p_w(1-p_c)+\frac{1}{2}p_wp_c.
\eeq
Combining these expressions we have that the winning probability is
\beq
p=1-\frac{1}{2}p_0-(1-p_0)p_1.
\eeq
In Figs. \ref{Fig: ExpEta}, \ref{Fig: ExpNu} and \ref{Fig: ExpEtaNu} we show the effect of loss and limited visibility on the winning probability of a (1-out-of-2) QRG with $n=4$ and a (1-out-of-3) QRG with $n=6$. The curves for the winning probability apply to both $k=2$ and $k=3$, since they are almost identical and indistinguishable in a single plot. As it can be seen from the figures, both limited visibility and efficiency decrease the winning probability, but even in the presence of realistic imperfections, it is possible to beat the bound $p>(1+\epsilon)/2$ of Ref. \cite{georgiou2015new} for the case $k=2$ and to achieve a significant gap compared to the cheating probability in the case $k=3$. However, this is no longer true for larger values of the imperfections. 

Thus, in order to experimentally demonstrate a QRG that can be used as a building-block in cryptographic applications, it will be necessary either to use very efficient detectors, low channel loss and high visibility interference, or to move to larger values of $k$ where the cheating probability decreases exponentially while the winning probability remains mostly unchanged. Moving to larger values of $k$ is something that can be done straightforwardly in our approach.

\section{Discussion}
We have given a general method of constructing (1-out-of-k) quantum retrieval games (QRGs) and we have shown that their cheating probability decreases exponentially in $k$ while the winning probability, in the ideal case, remains unchanged. Inspired by this, we have defined new QRGs based on coherent states of light. These QRGs can be implemented using only sequences of phase-modulated coherent states and linear optics with active switching. Such an implementation may be challenging, particularly in terms of achieving fast switching and an adjustable path difference in the interferometer, but it should be possible to realize with current technology, even for large values of $k$. 

While (1-out-of-k) QRGs may be constructed from (1-out-of-2) QRGs, this requires winning probabilities very close to 1 and small error probabilities, which may not be possible to achieve simultaneously in a practical setting. Similarly, using many (1-out-of-2) QRGs to build (1-out-of-2) QRGs with better parameters requires meeting conditions that may not be attainable in the presence of imperfections. Thus, our direct construction is probably a more desirable path as one can obtain large differences between the winning and error probabilities even in the presence of imperfections. 

From the point of view of applications, QRGs have already been used as building-blocks for quantum money schemes and therefore our results bring the experimental realization of such schemes closer to fruition. Additionally and perhaps most importantly, they constitute a new tool in the arsenal of the practical quantum cryptographer, which may prove useful in situations where we want participants to have access to certain information of their choice but not to all available information. Such situations arise for example in two-party cryptography and quantum signature schemes, where our results may be useful to build new practical protocols.

J.M. Arrazola would like to thank A. Ignjatovic and I. Kerenidis for useful discussions. M. Karasamanis is thankful for the hospitality of the Institute for Quantum Computing during a summer research term, in the course of which this research was performed. We acknowledge support from the Mike and Ophelia Lazaridis Fellowship, IQC Summer URA, NSERC Discovery Grant, Industry Canada, Army Research Laboratory W911NF-15-2-0061, Singapore Ministry of Education (partly through the Academic Research Fund Tier 3 MOE2012-T3-1-009) and the National Research Foundation of Singapore, Prime Minister’s Office, under the Research Centres of Excellence programme.
\bibliography{Bibliography}
\bibliographystyle{apsrev}

\end{document}